# TorchGWAS 1.0: GPU-accelerated GWAS at scale

Xingzhong Zhao[1†], Ziqian Xie[1†], Sheikh Muhammad Saiful Islam[1], Tian Xia[1], Degui Zhi[1]

[1]Department of Bioinformatics and Systems Medicine, McWilliams School of Biomedical Informatics, The University of Texas Health Science Center at Houston, Houston, Texas, United States of America

[†] Xingzhong Zhao and Ziqian Xie contributed equally to this work.

## Abstract

**Motivation:** Imaging, molecular and machine-learning workflows can generate thousands of quantitative phenotypes in one cohort. Repeating genome scans trait by trait creates a computational and output bottleneck.

**Results:** TorchGWAS is a GPU-accelerated framework that uses batched operations for high-throughput, covariate-adjusted linear association testing across large panels of quantitative phenotypes. Across 500,036 allele-harmonized tests, TorchGWAS $t$ statistics agreed with PLINK 2.0 (Pearson $1 - r = 1.3 \times 10^{-12}$). On an NVIDIA H100 80-GB GPU with a 48-core Intel Xeon Gold 6442Y host and measured disk read and write rates of 5.98 and 1.49 GB/s, respectively, median end-to-end times for 4.57 billion associations (8,931,083 variants by 512 phenotypes in 35,365 samples) were 28.46 s for BED, 29.13 s for hard-call PGEN, 51.48 s for BGEN and 58.95 s for dosage PGEN. These measurements included writing 36.7 GB of binary summary statistics. These results demonstrate TorchGWAS's efficiency for parallel fixed-effect association screening at scale.

**Availability and implementation:** TorchGWAS is implemented in Python and distributed as a documented source repository at https://github.com/ZhiGroup/TorchGWAS.

**Keywords:** GWAS, GPU computing, imaging genetics, deep phenotyping

## 1 Introduction

Genome-wide association studies have become a standard approach for identifying genetic influences on complex traits.[1] As cohorts grow larger and phenotyping becomes increasingly high-dimensional, the computational burden of GWAS grows with the total number of associations to be evaluated. This is especially apparent in resources such as UK Biobank and in modern machine-learning pipelines, where large numbers of candidate traits can be derived from the same individuals.[2] In these settings, repeatedly scanning the same genotype matrix creates substantial redundant computation and data movement.

Existing high-performance association tools address related problems but differ from TorchGWAS in their statistical scope or computational design. tensorQTL[3] is the closest computational comparator, but its trans-QTL implementation does not pipeline genotype loading, GPU computation and result writing, leaving these stages sequential at the workflow level. Its optional dense-output mode materializes the complete summary-statistic matrices in memory rather than streaming chunked results to disk in the current implementation. PLINK 2.0[4,5] provides highly optimized multithreaded CPU-based linear association testing and can process multiple phenotypes in a single run. fastGWA[6] supports efficient CPU-based linear and mixed-model analyses organized around trait-wise scans, whereas SAIGE-GPU[7] focuses on GPU-accelerated mixed-model analyses of quantitative and binary traits. Separately, permGWAS[8] targets GPU-supported permutation-based inference and rMVP[9] emphasizes CPU-level parallelization. Together, these tools address important computational bottlenecks in GWAS, but none combines association-level GPU parallelism with an end-to-end pipeline for dense fixed-effect screening at scale.

TorchGWAS implements covariate-adjusted linear GWAS using batched matrix operations over streamed genotype blocks and quantitative phenotypes. It performs conventional fixed-effect OLS within an end-to-end pipeline that overlaps genotype input, GPU computation and summary-statistic output, enabling efficient association testing at scale. This design supports rapid genome-wide screening in unrelated or relatedness-filtered samples. The overall workflow is showed in Fig. 1, with full protocols and diagnostics provided in the Supplementary Information.

## 2 Methods

### 2.1 Exact covariate-adjusted linear association

TorchGWAS performs covariate-adjusted fixed-effect OLS for each variant–phenotype pair. Covariate effects are removed from both the phenotype and genotype. Genotype residualization is evaluated through covariate-projection products without explicitly materializing a residualized genotype matrix. For each variant and each phenotype, samples with missing genotype or phenotype values are effectively excluded, so the analyzed sample size and residual degrees of freedom vary across variant–phenotype pairs. Each pair uses its own degrees of freedom to estimate the residual variance and standard error. To avoid numerical underflow in the extreme tail, TorchGWAS calculates $-\log_{10}P$ directly from the $t$ statistic and its degrees of freedom in log space using an incomplete beta function implemented in PyTorch. The resulting $P$ values are two-sided and are directly comparable with PLINK 2.0 when samples, allele coding, phenotype scaling and covariates are aligned. Full details are provided in Supplementary Methods S1 and Supplementary Table S2.

**2.2 Streaming GPU pipeline**

TorchGWAS accepts BED, PGEN and BGEN genotype files and aligns phenotype and covariate tables by sample identifier. It processes successive genotype chunks through a bounded pipeline that overlaps genotype loading and decoding, transfer to the GPU, association testing, result transfer and disk writing (Fig. 1). While the GPU evaluates one chunk, the next chunk is prepared in host memory and results from preceding chunks are written in the background. Reusable buffers and ordered queues preserve result order without holding the complete genotype or summary-statistic matrices in memory.

TorchGWAS supports association calculations in either float32 or float64 precision, with float64 mainly for validation or analysis on hardware with suitable double-precision performance. When chunk size is not specified, TorchGWAS uses a conservative memory-based heuristic to select a chunk size, accounting for free GPU memory and the estimated buffer footprint. The phenotype matrix remains resident on the GPU, so panels that exceed available memory must currently be divided into non-overlapping phenotype groups. Accelerated execution requires a CUDA-capable GPU. TorchGWAS was tested on compute capability 7.5 or newer, though compute capability 6.1 should also be supported when used with a compatible PyTorch/CUDA installation and kernels compiled for that architecture. Because variants are processed in chunks, peak GPU memory is independent of the total number of variants and depends primarily on the numbers of samples and phenotypes, the genotype representation, pipeline depth and variant chunk size. In the full-genome BED benchmarks with 35,365 samples, peak GPU allocation ranged from 0.8 to 1.9 GB across 1–2,048 phenotypes (Supplementary Methods S2).

Genotype loading and decoding, data transfer, GPU association testing and result writing were timed separately during pipeline development. These stage-level measurements were used to select chunk, buffer and queue configurations that reduce idle time and expose the limiting stage for each genotype format. Within each chunk, batched genotype–phenotype and covariate-projection operations calculate all $B \times K$ association statistics, where $B$ is the number of variants in the chunk and $K$ is the number of phenotypes. The statistics are streamed to tiled binary output (Supplementary Methods S3).

**2.3 Validation and benchmarking**

Because TorchGWAS fits a fixed-effect model, close relatives should be excluded before analysis. The accompanying workflow uses PLINK 2.0 --king-cutoff 0.044194 to exclude one individual from each pair estimated to be related at the third-degree level or closer.

Numerical concordance was evaluated across TorchGWAS, direct full OLS, and PLINK 2.0 using identical complete-case inputs and genotype-derived principal components as covariates**.** PLINK 2.0 was used as the primary concordance comparator because its linear association model jointly fits genotype and covariates, matching TorchGWAS's fixed-effect OLS specification. Null phenotypes were then generated from $Y = C\Gamma + E$ (where $C$ contains covariates including genotype-derived principal components, $\Gamma$ specifies their effects on the simulated phenotypes, and $E$ is random noise) using real genotype data to preserve linkage disequilibrium and structure. The null experiment comprised 3.2 million tests and varied the phenotype variance explained by three genotype PCs from 0% to 40%. Power was evaluated at $P < 5 \times 10^{-8}$ across 480 conditions using real genotypes from 5,000 samples, four minor-allele-frequency targets, five effect sizes, three SNP-PC correlation categories and single- and multiple-causal-variant architectures.

Runtime benchmarking was conducted across multiple problem sizes, phenotype counts, hardware platforms, comparison tools, genotype formats and output modes, using end-to-end wall-clock time from genotype loading through completion of summary-statistic writing. Each TorchGWAS timing used a fresh process, and PLINK 2.0 benchmarks requested 48 threads. Unless otherwise specified, benchmarks used float32 association arithmetic. Comparator benchmarks used PLINK 2.0 v2.00a7.5LM (10 September 2026) and GCTA v1.95.3 (official Linux AppImage) for fastGWA-lr, which refers specifically to GCTA's fixed-effect linear-regression mode. In the tested GCTA binary, the fastGWA association loop used a single CPU thread, and runtime did not materially change when the requested thread count was increased from 4 to 48. Resource-based runtime predictors were developed for TorchGWAS, PLINK 2.0 and fastGWA-lr by combining source-level operation and byte counts with independently measured hardware rates. They were used to estimate the workload scales at which TorchGWAS becomes faster than the comparator tools on different hardware platforms. The predictions were then evaluated through end-to-end experiments across the selected systems and workloads (Supplementary Methods S3–S4).

## 3 Results

### 3.1 Statistical correctness and calibration

Using float64 precision, TorchGWAS reproduced direct full OLS across all 192 tests, with a maximum absolute *P*-value difference of $5.66 \times 10^{-15}$ (Supplementary Fig. S1). The production-oriented float32 mode retained identical effect directions and genome-wide significance classifications in this test, with maximum absolute differences in the $t$ statistic and $-\log_{10}P$ of $9.24 \times 10^{-5}$ and $1.17 \times 10^{-4}$, respectively. PLINK 2.0 also agreed with direct full OLS after allele and scale alignment (Supplementary Table S2).

In a separate audit across three-dataset, 500,036 allele-harmonized tests were common to TorchGWAS and PLINK 2.0. TorchGWAS and PLINK 2.0 showed essentially exact agreement in $t$ statistics (Pearson $1 - r = 1.3 \times 10^{-12}$).

Across four structured-null scenarios (Supplementary information), TorchGWAS showed well-calibrated type I error rates at the evaluated thresholds and matched PLINK 2.0 false-positive counts. Omitting PCs led to increasing inflation as population structure explained more phenotype variance, whereas phenotype-only residualization was mildly conservative (Supplementary Fig. S2 and Table S3).

In the power experiment, TorchGWAS, direct full OLS and PLINK 2.0 produced identical detection counts across all 480 conditions (Supplementary Figs. S3 and Table S4).

### 3.2 Computational behavior

On an NVIDIA H100 80-GB GPU with a 48-core Intel Xeon Gold 6442Y host, TorchGWAS evaluated 4.57 billion associations across 8,931,083 variants, 512 phenotypes and 35,365 samples. With cold genotype and metadata caches, median end-to-end times were 28.46 s for BED, 29.13 s for hard-call PGEN, 51.48 s for BGEN and 58.95 s for dosage PGEN. These measurements included genotype loading and decoding, association testing and writing 36.7 GB of binary summary statistics (Supplementary Table S5).

Controlled experiments separated association scaling from input and output costs. With 5,000 samples and 100,000 variants, TorchGWAS computed 500 million complete association statistics across 5,000 phenotypes in 22.6 s. Increasing the phenotype count from one to 5,000 increased the number of tests 5,000-fold but increased compute time by only 15.7-fold. In a separate full-genome BED experiment with output disabled, scan time increased from 14.21 s for one phenotype to 14.90 s for 512 phenotypes and 18.09 s for 1,024 phenotypes, consistent with pipeline overlap until GPU computation became limiting.

Stage-level profiling separated genotype fetching and decoding, GPU association testing and result transfer. In a separate full-scale, no-output profile of the 512-phenotype hard-call PGEN scan, the overlapping stage clocks recorded 16.84 s for genotype fetching and decoding, 9.05 s for GPU association testing and 0.81 s for result transfer within an 18.98-s scan. Because these stages execute concurrently, their times are not additive. The profile confirms substantial overlap and identifies genotype preparation as the longest stage.

We next compared TorchGWAS with tensorQTL, PLINK 2.0 and fastGWA-lr across workload sizes and phenotype counts. Here, $M$, $N$ and $K$ denote the numbers of variants, samples and phenotypes, respectively. In matched no-output H100 experiments with 8,192 samples and 65,536 hard-call PGEN variants, TorchGWAS was 1.31–1.56-fold faster than tensorQTL v1.0.10 in trans-QTL mode at every evaluated $K$ (1–32). For the CPU comparators, measured single-phenotype comparisons across varying sample and variant counts placed the fastGWA-lr crossovers on the order of $MN \approx 10^9$ and those for PLINK 2.0 with 48 threads on the order of $MN \approx 10^{10}$. Resource-based runtime predictors combining source-level operation and byte counts with independently measured hardware rates reproduced these general transitions and extended the comparison across hardware platforms (Fig. 2b). Together with the sublinear increase in TorchGWAS runtime for moderate $K$, these results support $KMN \approx 10^{10}$ (for example, 10,000 samples × 1 million variants × 1 phenotype or 2,000 samples × 500,000 variants × 10 phenotypes) as an approximate workload scale above which TorchGWAS was faster than the comparators in the evaluated settings.

## 4 Discussion

TorchGWAS addresses an increasingly important need for efficient GWAS at scale. By pipelining genotype loading, GPU computation and result writing while evaluating associations in batches, it scales efficiently across sample, variant and phenotype dimensions. TorchGWAS outperformed the CPU comparators even in single-phenotype analyses at sufficiently large sample and variant counts, while supporting efficient analysis of large phenotype panels. The empirical crossover depends on cohort size, variant count, phenotype count, storage and hardware (Table S6). End-to-end performance can become limited by genotype decoding and result writing rather than matrix multiplication alone.

TorchGWAS is a fixed-effect association framework rather than a complete replacement for established GWAS software. Its association statistics can be used directly when fixed-effect OLS is appropriate. The current release supports linear association testing and a preprocessing workflow for excluding related individuals, but not generalized linear models or mixed-model inference. Established mixed-model methods remain necessary when retained relatedness or a genomic random effect must be modeled.

TorchGWAS produces summary statistics for each retained variant–phenotype pair, allowing downstream prioritization by criteria such as the minimum *P* value or the number of independent significant loci. LD clumping and multiple-testing correction across phenotypes are not performed automatically and should be applied in downstream analyses as appropriate.

## Availability and implementation

TorchGWAS is implemented in Python and is distributed as a documented source repository at https://github.com/ZhiGroup/TorchGWAS. The repository contains packaged source code, installation instructions, tutorials, benchmark scripts, toy examples, and templates for local real-data workflows.

## Funding

This work was supported by the National Institute on Aging under award U01AG070112.

## Conflict of interest

None declared.

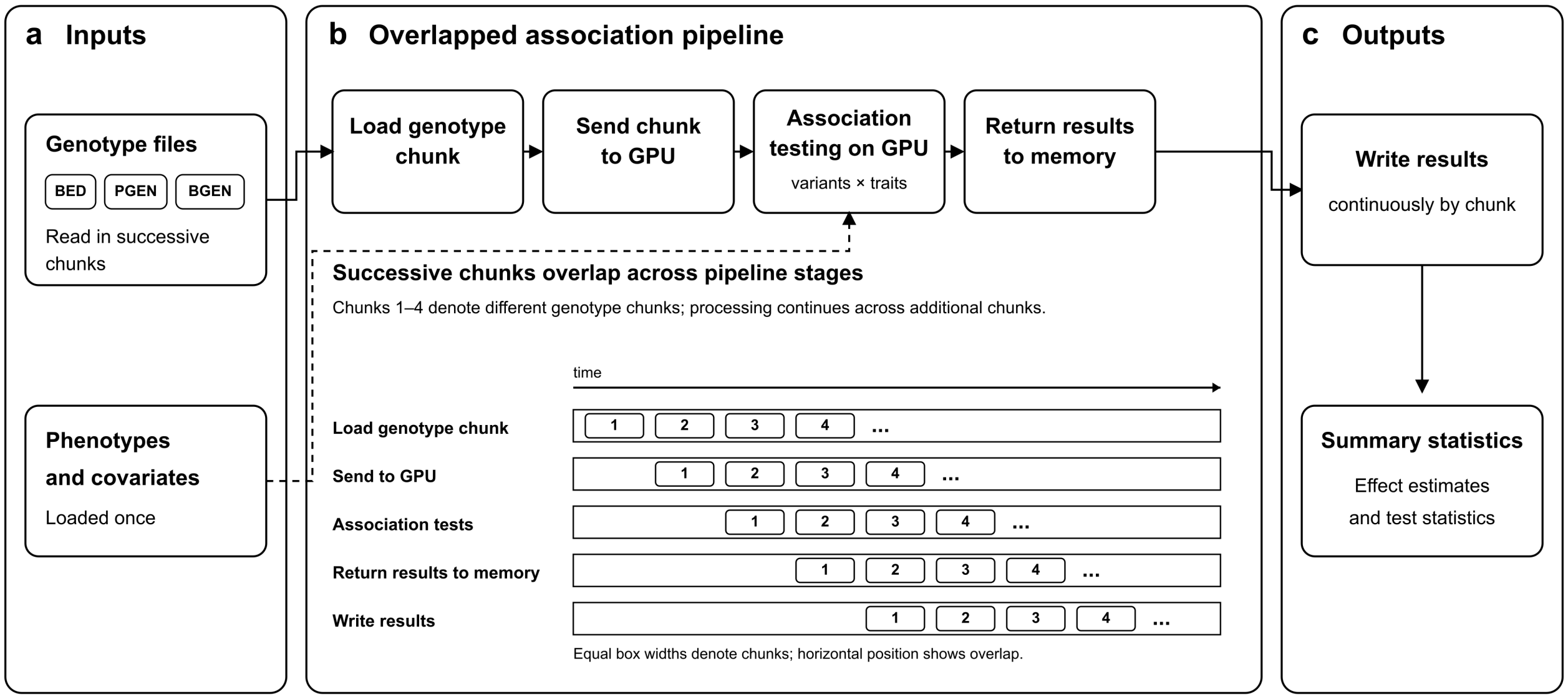

**Fig. 1. Overview of the TorchGWAS pipeline.** TorchGWAS processes genotype data in successive chunks and uses the GPU to test many associations in parallel. The pipeline overlaps data loading, association testing and result writing to improve efficiency. Writing results by chunk avoids retaining the complete result matrix in memory.

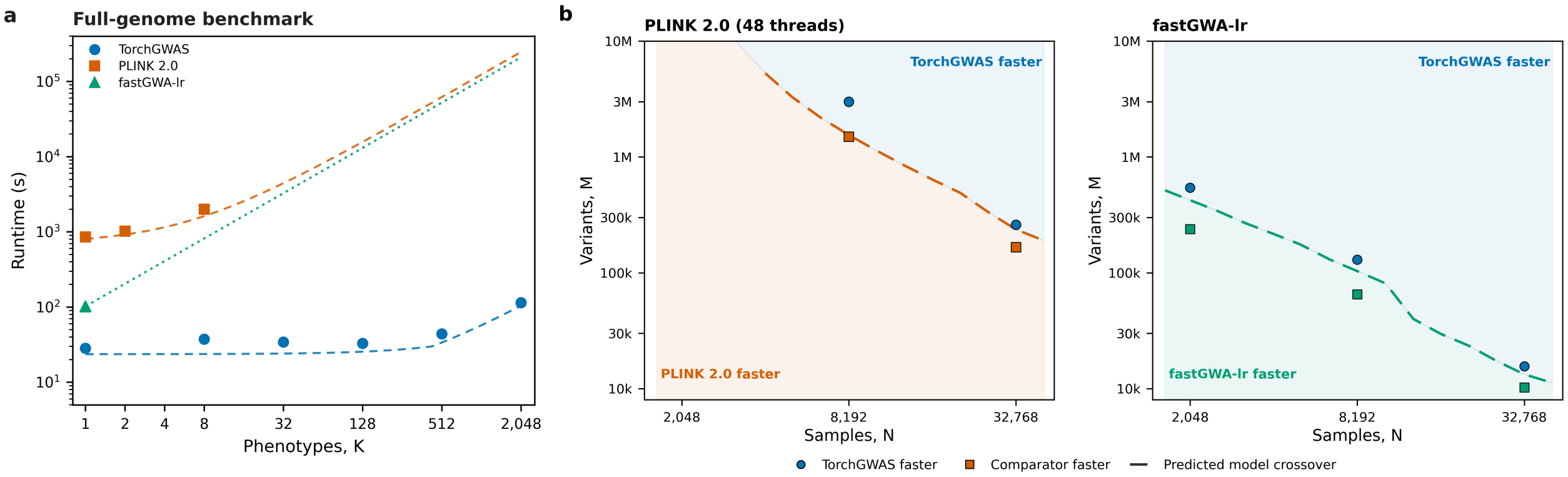

**Fig. 2. End-to-end runtime scaling and predicted crossover regimes.** (a) Full-genome runtime on an NVIDIA H100 80-GB GPU for 8,931,083 variants in 35,365 samples as the number of phenotypes (K) increased. Symbols show measured end-to-end runtimes, including result writing. Dashed curves show estimates from the TorchGWAS and PLINK 2.0 runtime predictors, whereas the dotted fastGWA-lr curve is a serial projection from its measured K = 1 runtime. The runtime predictor curves were not fitted to the plotted measurements. (b) Single-phenotype runtime crossover across sample size (N) and variant count (M) for PLINK 2.0 (48 threads) and fastGWA-lr. Filled circles indicate measured configurations in which TorchGWAS was faster, filled squares indicate configurations in which the comparator was faster. Dashed curves show the equal-runtime boundaries estimated by the runtime predictors, and background shading indicates the method predicted to be faster on either side. Each configuration in panel b was measured twice. For each crossover with measurements on both sides, the nearest available bracketing points are shown.